\documentclass[english,aps,prper,reprint,showpacs,titlepage,longbibliography,nofootinbib,superscriptaddress]{revtex4-2}

\usepackage[T1]{fontenc}	
\usepackage[latin9]{inputenc}	
\usepackage{geometry}		
\usepackage{graphicx}
\usepackage[above,below]{placeins}	
\usepackage{times}
\usepackage{multirow}
\usepackage[dvipsnames]{xcolor}
\usepackage{subcaption}
\usepackage{titlesec}
\usepackage{ifthen}
\usepackage{makecell}
\newenvironment{iquote}
    {\vspace{-.33\baselineskip}\itshape\list{}{\leftmargin=0.15in\rightmargin=0.15in}%
    \item\relax}
    {\endlist\vspace{-.33\baselineskip}}

\newcommand{\ket}[1]{|#1\rangle}
\newcommand{\bra}[1]{\langle#1|}

\newcommand{\fracroot}[2]{\ifthenelse{#1=1}{\frac{1}{\sqrt{#2}}}{\sqrt{\frac{#1}{#2}}}}

\usepackage{makecell}

\newcommand{\who}[1]{\textbf{#1\emph{:}}}

\newif\ifshowtimestamp
\showtimestampfalse

\newcommand{\timestamp}[1]{
    \ifshowtimestamp{\textcolor{red}{Timestamp: #1}}\fi}

\usepackage{hyperref}  
\hypersetup{colorlinks=true,urlcolor=blue,citecolor=blue,linkcolor=blue}   
\usepackage{enumitem}        
\setlist{nosep}                 

\usepackage{mathtools}
\usepackage{amssymb}
\usepackage{amsmath}
\usepackage{amsthm}
\usepackage[font=small]{caption}
\usepackage{subcaption}
\usepackage{bbold}
\usepackage{comment}
\usepackage{tabularx}
\usepackage{array}
\usepackage{censor}

\usepackage[dvipsnames]{xcolor}
\usepackage{footmisc}

\usepackage[sort&compress]{natbib}

\begin{document}

\begin{titlepage}

\title{The interdisciplinary quantum information classroom:\\ Themes from a survey of quantum information science instructors}
\author{Josephine C. Meyer}
\affiliation{Department of Physics, University of Colorado - Boulder, Boulder, CO 80309}
\author{Gina Passante}
\affiliation{Department of Physics, California State University - Fullerton, Fullerton, CA 92831}
\author{Steven J. Pollock}
\author{Bethany R. Wilcox}
\affiliation{Department of Physics, University of Colorado - Boulder, Boulder, CO 80309}
\begin{abstract}
Interdisciplinary introduction to quantum information science (QIS) courses are proliferating at universities across the US, but the experiences of instructors in these courses have remained largely unexplored in the discipline-based education research (DBER) communities. Here, we address this gap by reporting on the findings of a survey of instructors teaching introduction to QIS courses at institutions across the US, primarily at the undergraduate or hybrid undergraduate/graduate level, as well as follow-up focus interviews with six individual instructors. Key themes from this analysis include challenges and opportunities associated with the diversity of instructor and student backgrounds, student difficulties with the mathematical formalism (especially though not exclusively with linear algebra), and the need for better textbooks and curricular materials. We also find that while course topics are ostensibly similar, each course is crafted by its instructor to tell a different story about QIS and to uniquely balance goals such as accessibility and academic rigor, such that no canonical introduction to QIS course emerges from our dataset. We discuss the implications of this finding with regard to the benefits and risks associated with standardization of curricula as QIS coursework matures.
\end{abstract}

\maketitle
\end{titlepage}

\section{Introduction and Motivation}

Quantum information science (QIS) coursework has, until recently, been restricted almost entirely to theoretical Ph.D.-level coursework in physics. However, fueled by rapid growth in demand for jobs in quantum computing in the private sector as well as funding from the National Quantum Initiative Act of 2018, interdisciplinary QIS courses at the undergraduate or combined undergraduate/graduate level have begun to proliferate across US universities \cite{Fox:2020}. Such courses cover topics as wide-ranging as quantum algorithms, cryptography, quantum computer programming, laboratory work, and hardware design. Due to their relative novelty, little research has been done yet on such courses by the physics education research (PER) community and similarly situated discipline-based education research (DBER) communities. Recognizing this need, 32 scientists and professionals in QIS and adjacent fields signed an open letter calling for, among other provisions, the early involvement of education experts in curriculum development \cite{Aiello:2021}.

QIS education represents an intriguing but challenging frontier in DBER communities for two reasons. First, QIS courses are often interdisciplinary across physics, computer science (CS), math, and/or electrical engineering departments \cite{Cervantes:2021}. Interdisciplinary courses present a particular challenge to DBER, since incoming students' backgrounds
often vary substantially within and across courses. At the same time, the relative infancy of QIS courses presents an opportunity to influence instruction from the beginning.

In previous work, our team used publicly-available course catalog information to compile a list of introductory QIS courses at US institutions and probed the nature and distribution of these courses \cite{Cervantes:2021}. 
Our work here seeks to provide a complementary perspective, drawing from instructor survey responses and targeted follow-up interviews to investigate QIS education from the instructor perspective. Through analysis of the survey data and case studies, we paint a picture of the diversity of existing QIS courses and the challenges, opportunities, and needs that come with teaching in such an interdisciplinary and rapidly-growing field of study.
More broadly, while QIS is still a relatively niche field, our findings have implications in all the interdisciplinary fields that QIS touches, with the potential to catalyze work in fields such as CS that have historically lagged behind physics in terms of DBER.

We focus on three research questions:

\begin{enumerate}
    \item Are the academic backgrounds of students in introductory QIS courses homogeneous or heterogeneous (both within and across courses)? How do these student backgrounds affect the teaching and learning of QIS?
    \item Are the disciplinary backgrounds of faculty teaching introductory QIS courses homogeneous or heterogeneous? How do faculty's personal backgrounds affect how introduction to QIS courses are taught?
    \item What are some of the challenges instructors face when teaching introduction to QIS courses, and how might DBER efforts help address these challenges?
\end{enumerate}

\section{Background}

\subsection{A Brief Overview of QIS Education}

In this section, we offer a brief overview of the essential similarities and differences between QIS and traditional quantum courses, as well as a brief primer on core topics in QIS.  Readers familiar with the field of QIS can skip to Sec.~\ref{sec:prior}.  

The National Strategic Overview for Quantum Information Science identifies four key areas of fundamental science under the umbrella of QIS: quantum sensing, quantum computation, quantum networking, and broader scientific advances enabled by advances in quantum theory and devices \cite{SCQIS:2018}. Of these, quantum computing and associated topics in quantum communication/networking tend to be the dominant areas discussed in the courses we analyze in this paper.

Classical computers process information using bits that are deterministically either 0 or 1. A quantum computer or communicator obeys similar principles, except the classical bit is replaced with a quantum bit, or "qubit," an effective spin-1/2 system where the state of each qubit can be any normalized linear superposition of the basis states $\ket{0}$ and $\ket{1}$\footnote{$\ket{0}$ and $\ket{1}$ are vectors or ``kets'' in an abstract vector space known as Hilbert space in quantum mechanics.} \cite{Mermin:2003}. 
Clever protocols can utilize the quantum nature of qubits to speed computation. Notable examples include Shor's algorithm \cite{Shor:1994} for factoring integers in polynomial time\footnote{In CS, time complexity of algorithms is measured by how the time (or equivalently, number of steps) required to solve a problem scales with input size. Algorithms are said to run in ``polynomial time'' if the time it takes to run an algorithm scales at worst polynomially with the input size. Algorithms whose time complexity scales superpolynomially (e.g.\ exponentially) with the input size are, by contrast, generally intractable for large inputs. Shor's quantum algorithm factors integers in polynomial time, whereas the best known classical algorithms run in exponential time.} and Grover's search algorithm \cite{Grover:1996}. Whereas the information encoded in a system of classical bits scales as $\mathcal{O}(N)$ where $N$ is the number of bits, the information encoded in a quantum state scales with the dimension of the Hilbert space of the qubits, that is as $\mathcal{O}(2^N)$. Likewise, the laws of quantum mechanics can be exploited in communication protocols to achieve superdense coding \cite{Bennett:1992} or secure quantum key distribution \cite{Bennett:1984}.

CS as a field owes much of its sucess to the ability to operate at a higher level of abstraction than individual 0s and 1s. Likewise, the focus of today's QIS courses differ, sometimes subtly and sometimes substantially, from that of a traditional quantum mechanics course. Depending on the goals of the course, the physical implementation of the qubits and logic gates may be considered a question of relatively minor importance. Applications such as quantum algorithms require relatively little knowledge about how a qubit is constructed but a sophisticated understanding of how a qubit \textit{behaves} and the underlying mathematical formalism \cite{Mermin:2003,Singh:2006}.

On the other hand, certain fundamental quantum mechanics principles are of particular importance in QIS precisely because they relate to the way information is encoded in a quantum state - most notably the principles of superposition and entanglement. \textbf{Superposition} refers to the fact that the state of any qubit can be an arbitrary linear combination of basis states (such as the computational basis states $|0\rangle$ and $|1\rangle$), whereas a classical bit can only take on the discrete values 0 and 1. \textbf{Entanglement} refers to the phenomenon in which the quantum state of one qubit cannot be written independently of the state of another qubit(s); as a result, measuring one qubit will irreversibly change the state of the other qubit(s). Until the measurement is made, information is correlated across qubits spaced arbitrarily far apart. Indeed, Seegerer, Michaeli, and Romeike \cite{Seegerer:2021} identify superposition and entanglement as two of the five core ideas of quantum computing education emergent from analysis of educational materials and expert interviews, alongside quantum computers themselves, quantum algorithms, and quantum cryptography.

\subsection{Prior Work}\label{sec:prior}

Though limited in number, publications by DBER researchers addressing topics of QIS education date back to the mid-2000s, often drawing from research in the context of traditional quantum mechanics courses. In 2003, Mermin \cite{Mermin:2003} proposed a conceptual framework for teaching essential quantum mechanics to computer scientists in the context of quantum computing, arguing that it was possible to teach computer scientists the necessary quantum mechanics without the structure of a typical physics degree. A year later, Grau \cite{Grau:2004} proposed a similar framework for introducing computer scientists and engineering students to quantum mechanics through quantum cryptography.

PER researchers have subsequently identified student difficulties in quantum mechanics classes particularly relevant to QIS education \cite{Singh:2006, Singh:2009, Zhu:2012, Kohnle:2015, Passante:2015, Singh:2015, Wan:2019, Li:2021}. Marshman and Singh \cite{Marshman:2015} contextualize student difficulties in quantum mechanics by noting that students enter quantum mechanics courses with different backgrounds and levels of preparation, and that the process of learning quantum mechanics requires a Kuhnian \cite{Kuhn:1962} ``paradigm shift'' away from the `realist' worldview of classical mechanics that often takes place in an incomplete or piecemeal fashion, an analysis that likely carries over to the context of the QIS classroom.
Recent work from our group has built upon these foundations to study student reasoning in quantum computing contexts specifically \cite{Meyer:2021}.

PER researchers have also developed curricular materials to improve student understanding of QIS topics, though with limited exceptions (e.g.\ \cite{Satanassi:2022}) these have generally been designed and evaluated in the context of traditional quantum mechanics courses \cite{DeVore:2014, DeVore:2020, Kohnle:2017} rather than true interdisciplinary QIS contexts. Researchers in engineering and CS education have engaged in similar efforts in more interdisciplinary QIS contexts \cite{Zapirain:2019, Salehi:2021}. Recent work has begun to create a roadmap for comprehensive QIS education at the undergraduate level and beyond \cite{Perron:2021, Asfaw:2021}.

\section{Methodology}

We gathered instructor perspectives using a two-step approach to data collection. First, a survey was distributed to QIS instructors across US institutions in order to gain insight into the scope and nature of introductory QIS education. Thematic analysis of initial survey responses were used to craft follow-up interviews with a subset of six faculty in order to explicate and interpret the themes emergent from the survey in response to our research questions.

\subsection{Faculty Survey}

\subsubsection{Survey Distribution and Design}

We developed an online survey drawing on prior experience with similar instructor surveys\footnote{Survey available upon request from lead author.} \cite{Rainey:2019}. We distributed the survey to faculty identified as potentially teaching or having taught courses containing significant QIS content. Emails were sent to 116 faculty with an invitation to participate in the survey in March-April 2021. Most recipients (109) were identified through the course catalog analysis discussed in Ref.~\cite{Cervantes:2021}, representing courses at 74 institutions.
Additionally, faculty teaching courses that did not meet the strict definition of "introductory QIS course" given in Ref.~\cite{Cervantes:2021} but whose courses were identified as potentially containing QIS content were also included in the initial email. We also sent the email to 7 faculty known to be QIS instructors who were not on the list produced by the search discussed in Ref.~\cite{Cervantes:2021}, and we asked faculty who received the email to forward it to any other faculty who might be interested in the project. A link to the survey was also distributed at a QIS education conference in June 2021 focusing on undergraduate education. 

Faculty were asked to report basic demographic information about their course, such as the department(s) in which the course is listed, estimated cross-sections of the student population in the course by year and major, textbook used (if any), and any prerequisites or expected prior preparation in math, CS, or physics. We also asked faculty to identify specific student difficulties they observed and challenges they faced as instructors, as well as needs or requests for DBER work in the field of QIS.
Questions were phrased as multiple-choice or free-response as appropriate. Prerequisite courses and student difficulties were qualitatively coded according to discipline and typical undergraduate course sequencing. A few longer free-response questions were used to elicit faculty's broader opinions about issues pertinent to QIS education; transcripts from these questions were analyzed thematically.

We received 32 completed responses, 27 from the initial email link and 5 from the link distributed at the conference. We found it necessary to adopt a working definition of introductory QIS courses so as to screen courses with little QIS theory from the dataset, defining an introductory QIS course as a lecture course with a primary focus on theory or applications of QIS as defined by the four areas of fundamental science identified by the Subcommittee on Quantum Information Science \cite{SCQIS:2018}. Applying this definition, we discarded 4 responses from the dataset (2 courses focused primarily on materials science of quantum device fabrication and 2 traditional quantum mechanics courses with a brief QIS unit) giving a total of $N=28$ survey responses specific to introductory QIS courses.

One of the remaining 28 responses (Response \#23) corresponded to an undergraduate course in development. We retained this response in our analysis of instructor and institution demographics and for aspects of the course that had already been established (e.g.\ prerequisites and cross-listing). We excluded it from research questions more directly focused on student or instructor experiences.

The survey was targeted at instructors of undergraduate and hybrid undergraduate-graduate courses, though we also received a total of 5 responses from instructors of exclusively graduate-level courses. We opted to include these 5 graduate courses in our analysis, given that a true introduction to QIS course will likely begin at a similar place academically whether the primary audience is at the undergraduate and graduate level. However, given the small number of graduate-only courses in our dataset and the undergraduate-oriented focus of our survey, our conclusions should be understood as applying most directly to undergraduate and hybrid-level courses. For this reason, none of the graduate courses were selected for follow-up interviews, and quotations from survey responses corresponding to graduate courses are explicitly identified as such.

\subsubsection{Respondent and Institution Demographics}

A priority for our survey distribution was to ensure a diverse range of instructor backgrounds and institutions were represented. Institutions were classified according to the Carnegie Classification of Institutions of Higher Education \cite{carnegie} basic classification and institutional control (Table~\ref{tab:carnegie}), as well as by geographic location. Findings are largely consistent with the distribution of courses discussed in Ref.~\cite{Cervantes:2021}. As in Ref.~\cite{Cervantes:2021}, we find that private institutions and R1 research institutions are significantly overrepresented in our study compared to the cross-section of US institutions as a whole (Table~\ref{tab:carnegie}).
 While we find that a strong majority of responses (68\%) corresponded to doctoral-granting institutions, this rate is significantly lower than the 86\% reported in Ref.~\cite{Cervantes:2021}, an effect we attribute to the survey distribution's inherent bias in favor of undergraduate courses and the fact that beyond-intro QIS courses included in Ref.~\cite{Cervantes:2021} excluded from the scope of this survey are likely to be disproportionately graduate courses. We also asked faculty to optionally report their race and gender; the respondents overwhelmingly self-identified as Caucasian and male (Table~\ref{tab:demographics}).

Institutions were also classified by MSI status using the November 2020 NASA List of Minority Serving Institutions \cite{NASA:2020}. Two institutions were identified as both AANAPISI (Asian American, Native American, and Pacific-Islander Serving Institutions) and HSI (Hispanic Serving Institutions). A third institution was identified as HSI but not AANAPISI. No other MSI categories were represented among responses.

\begin{table}[]
    \centering
    \begin{tabular}{c c c c}
        \hline \hline
        \thead{Carnegie Classification} & \thead{Count} & \thead{Control} & \thead{Count}\\
        \Xhline{2\arrayrulewidth}
        Doctoral (R1) & 18 & Private & 16\\
        Doctoral/Professional & 1 & Public & 12\\
        Masters (M1) & 3 & &\\
        Baccalaureate & 6 && \\
        \Xhline{2\arrayrulewidth}
        Total & 28 & Total & 28 \\
        \hline \hline
    \end{tabular}
    \caption{Institutional classification and institutional control (public/private) for each of the 28 institutions represented in our analysis. All classifications courtesy of the Carnegie Classification of Institutions of Higher Learning \cite{carnegie}.}
    \label{tab:carnegie}
\end{table}

\begin{table}[]
    \centering
    \begin{tabular}{c c c c}
        \hline \hline
        \thead{Instructor Race} & \thead{Count} & \thead{Gender} & \thead{Count}\\
        \Xhline{2\arrayrulewidth}
        Caucasian & 21 & Man & 24\\
        Asian & 4 & Woman & 4\\
        Hispanic/Latinx & 2\\
        Black/African American & 1 & &\\
        Do not wish to disclose & 2 & &\\
        \Xhline{2\arrayrulewidth}
        Total & 28* & Total & 28 \\
        \hline \hline
    \end{tabular}
    \caption{Self-reported instructor race and gender. Race and gender categories with zero responses not shown. Racial categories based on US census categories. *Instructors were allowed to select more than one racial category.}
    \label{tab:demographics}
\end{table}

\subsection{Follow-Up Interviews}

Following preliminary analysis of the survey responses, a follow-up focus interview protocol was devised with the goal of better understanding the causes and implications of the trends observed in the survey with regard to our research questions. Interviewees were selected with the intent of ensuring a diversity of instructors background, institution type, and student cross-section, with a secondary goal of selecting respondents whose initial survey responses were particularly generative. We also preferentially selected faculty whose courses were crosslisted across multiple departments and/or course levels so as to better sample diverse student backgrounds. Graduate-only and freshman-only courses were excluded from the interviews since these courses are beyond the scope of our current curriculum-development efforts.

Given our selection methodology, it is likely that the 6 faculty interviewees disproportionately represent instructors who are particularly committed to QIS education and have strong opinions on the matter. Additionally, the 6 faculty who agreed to participate in the interviews all self-identify as Caucasian men (race and gender were not considered in the selection process for follow-up interviewees) and none teach at MSIs. Basic background information on each of the interviewees and their course is shown in Table~\ref{tab:interviewees}.

Faculty were asked to reflect on themes from the survey analysis: specifically, the diversity of student backgrounds in the course, the impact of the instructor's own background, concerns about student difficulties with math and especially linear algebra, and concerns about textbook quality. Faculty were also asked to provide their input on any other unique challenges specific to QIS courses, to describe their background and process for arriving at teaching the course, and how the course has evolved over time (if applicable). Finally, interviewees were asked to elaborate on the types of curricular and/or assessment materials they would like DBER researchers to ultimately develop.

The interviews were given in a semi-structured format and thus varied somewhat from faculty member to faculty member. Interviews ranged in length from approximately 50 minutes to 3 hours. Five of the 6 interviews were conducted over Zoom, and all interviews were recorded and transcribed using Otter.ai. Transcripts were hand-verified for accuracy as appropriate.

\begin{table*}[tb]
    \centering
    \begin{tabular}{c c c c c c c c}
        \hline \hline
         \thead{Course} & \thead{Pseudonym} & \thead{Home Dept} & \thead{Institution Type} & \thead{Course Level} & \thead{Listed Dept(s)} &  \thead{Since}  \\
         \Xhline{2\arrayrulewidth}
         \thead{A} & Albert & Computer Science & Private R1 & BFY Undergrad & Computer Science & 1999\\
         \thead{B} & Ben & Physics & Private R1 & Hybrid & Physics & 2020\\ 
         \thead{C} & Carl & Physics & Public R1 & BFY Undergrad & \makecell{Physics, Computer Science} & 2020\\
         \thead{D} & David & Physics & Public R1 & Hybrid & \makecell{Physics, Computer Science} & 2011\\
         \thead{E} & Edwin & Physics & Private R1 & Hybrid & \makecell{Physics, Computer Science} & 2000*\\
         \thead{F} & Franz & Math & \makecell{Private Baccalaureate} & BFY Undergrad & Computer Science & 2018\\
         \hline \hline
    \end{tabular}

    \caption{Demographic profiles of the six courses and their instructors for which follow-up interviews were conducted. Institution type is from the Carnegie Classification of Institutions of Higher Learning \cite{carnegie}; all information self-reported by instructors. ``Physics'' includes engineering physics. BFY indicates courses intended for an audience of beyond-first-year undergraduates.
    *Date approximate because course predates current instructor, who began offering it in 2008.}
    \label{tab:interviewees}
\end{table*}

\subsection{Limitations of This Study}

Our study's primary purpose is to better understand and interpret faculty's \textit{perceptions} of QIS teaching and learning, which -- though undoubtedly valuable -- are often mere windows into broader underlying trends and issues.
We caution against the conflation of the descriptive with the prescriptive:
instructors are experts on their own experience, not necessarily on curricular reform or student reasoning, for example. 

Our results should also be considered in the context of the inherent limitations associated with studying such a diverse and evolving field of coursework. Most importantly, our sample size of $N=28$ courses -- though reasonably consistent with our prior experience with faculty survey response rates \cite{Rainey:2019} -- is too small for strong statistical analysis of quantitative data. Though we find it useful to present some quantitative results from our survey data, we opt not to present claims of statistical significance, encouraging the reader to interpret such results not as quantitative claims that can be extrapolated to all QIS courses but rather as a snapshot of the diversity of QIS courses present in our responses at the time of the survey (spring 2021). Even with larger $N$, we would be hesitant to make statistical claims simply because the field of QIS education is evolving so rapidly that any such claims would likely become quickly outdated.

Finally, we acknowledge that, in part given the unequal distribution of QIS courses across US institutions and existing inequities within academia, our methodology fails to adequately capture the unique perspectives that faculty with marginalized identities and faculty teaching at non-R1 institutions bring to the table. Moreover, all four co-authors are physics education researchers and are trained as physicists, and we acknowledge that our positionality within the discipline of physics will influence our interpretation of our findings and our underlying research questions.

\section{Key Findings}

In this section, we discuss themes from our analysis of the survey and follow-up interview responses. Findings are grouped in order by research question (see above). Where possible, we summarize key claims in the first paragraph of each subsection and reserve subsequent paragraphs for justifying and explicating these claims.

\subsection{Diversity of Student Backgrounds and Its Effect on Instruction}

Our first research question sought to identify the variation in students' academic background within and across introduction to QIS course and what effects any such variation might have on instruction. We find that students in intro to QIS courses vary significantly in disciplinary background, preparation, motivation, and other axes that are likely to affect instruction. We also find that the diversity of student backgrounds presents both unusual opportunities and unique challenges for students and instructors in introductory QIS courses compared to the upper-division and graduate courses traditionally treated within PER.

Our goal in this section is not to paint a profile of the ``typical'' introductory QIS course, for if anything, our results suggest such a course may not exist (yet). Each introductory QIS course has a different intended and actual audience, and its students possess different motivations, prior knowledge, and disciplinary backgrounds. We highlight the interplay between design decisions in implementing a course and the audience of students the course attracts.

\subsubsection{Diversity of Student Academic Backgrounds}

We find that introductory QIS courses target students of diverse disciplinary cross-sections both within and across courses. We consider catalog listing and instructor estimates of the majors of students as indicators of the academic background for which the course is targeted.
Given the diverse academic backgrounds of students, we find unsurprisingly that instructors teaching introduction to QIS courses perceive conflict between the goals of accessibility across disciplines and academic rigor; instructors take a variety of approaches in balancing these objectives, which we discuss below.

\begin{figure}
    \centering
    \includegraphics[width = 0.5 \textwidth]{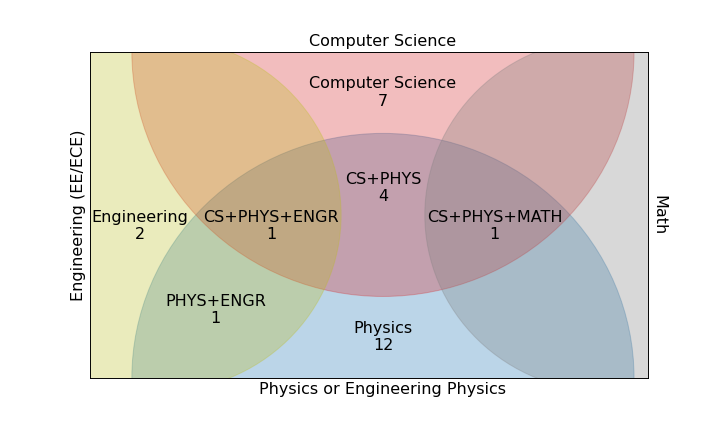}
    \caption{Venn Diagram illustrating the academic departments in which the 28 surveyed courses are (or are intended to be, in the case of Respondent \#23) listed or crosslisted. ``Physics'' includes engineering physics. Where course catalogs differentiate between subfields of engineering, courses classified here as ``Engineering'' were listed in either Electrical Engineering or Electrical and Computer Engineering.}
    \label{fig:crosslistings}
\end{figure}

Figure \ref{fig:crosslistings} shows the academic departments in which each of the 28 intro to QIS courses is located or intended to be located. Where institutional catalogs distinguish between subfields of engineering, all listings classified as ``Engineering'' correspond either to Electrical Engineering (EE) or Electrical and Computer Engineering (ECE). We find that the majority of introductory QIS courses represented in this survey are listed in physics and/or computer science departments, though listings in EE/ECE and math are also represented in the sample. Fully 25\% of the surveyed courses ($N=7$) are cross-listed in two or more departments, a strong indication that the instructor has an interdisciplinary audience in mind.

\begin{figure}
    \vspace*{-1cm}
    \hspace*{-1cm}\includegraphics[width = 0.6 \textwidth]{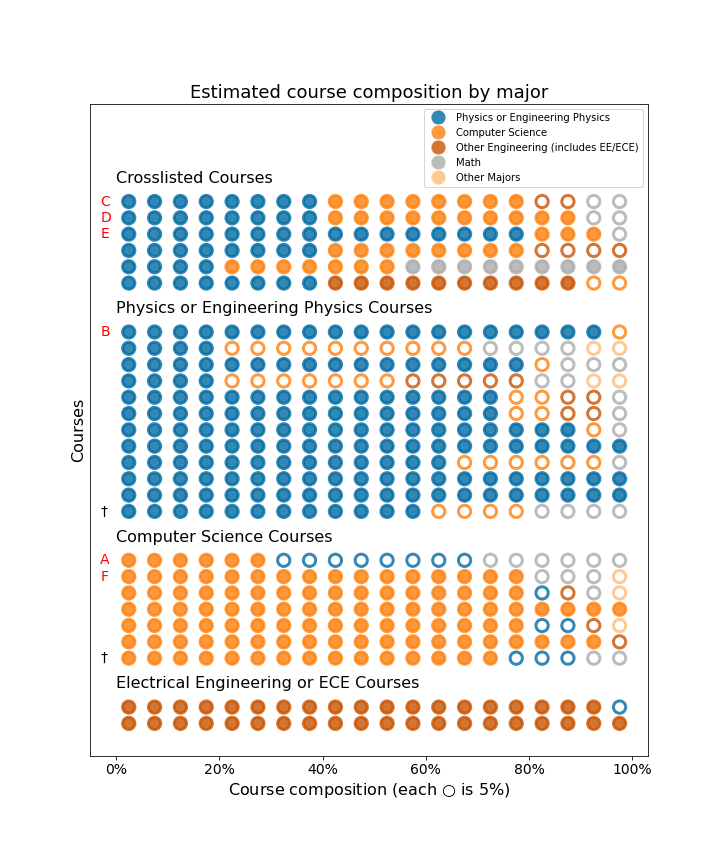}
    \vspace*{-1cm}
    \caption{Estimated enrollment profile of each of the 27 previously-taught courses. Each circle represents 5\% of students in the course. Solid circles represent students in a major that corresponds to one of the listed departments for the course. Hollow circles represent students in a major that does \textit{not} correspond to one of the listed departments for the course. Courses A-F that are the subject of follow-up interviews are labeled as such. Dagger ($\dagger$) indicates that instructor estimates were renormalized by our team to sum to 100\%.}
    \label{fig:enrollmentprofile}
\end{figure}

Given the uniquely interdisciplinary nature of QIS research, we anticipated that catalog listing may not reflect the actual diversity of academic backgrounds of students in introduction to QIS courses. Therefore, we also asked survey respondents to estimate the actual composition of their courses by major. Figure \ref{fig:enrollmentprofile} shows instructors' estimates of the distribution of majors for each of the 27 courses that had previously been offered. Enrollment profiles for each major are rounded to the nearest 5\% corresponding to the typical granularity of estimates provided by instructors\footnote{We observed that all but one faculty member gave enrollment estimates as integer multiples of 5\% or 10\%. Though we have no independent means to corroborate the reliability of instructor estimates, we take faculty's implicit consensus on the reliability of their own estimates as a reasonable bound on their overall reliability}. Though we observe a rough correlation between catalog listing and actual course composition, the picture is more complicated at the level of individual courses. One of the cross-listed courses is 80\% physics majors, for instance, while the majority of courses listed exclusively in physics or engineering physics (7/12) and computer science (5/7) had at least 20\% majors from outside the listed department.

From the survey responses and follow-up interviews, we learned that many courses incorporate a diverse group of students by design. This intentional decision is illustrated by Carl, whose course features a near-even split between physics and CS majors. Though Carl's background is in physics, he intentionally sought other perspectives from the beginning in recruiting co-instructors:

\begin{iquote}
    \who{Carl}
    We really wanted to work with someone in CS, too, because ... we didn't want there to be a CS version of the class, and a physics version of the class, and [an] ECE version of the class.
    \timestamp{4:30 verified}
\end{iquote}

Carl's course very intentionally appeals to a diverse audience of students. Whenever possible, each iteration of the course is taught by two or more faculty with different disciplinary and research backgrounds. The course is crosslisted in both the physics and CS departments. He is very explicit about his intent of designing the course for accessibility including across disciplines:

\begin{iquote}
    \who{Carl}
    Yes, our goal was [to] make this as accessible as we could.
    \timestamp{21:02 verified}
\end{iquote}

Ben approached the design of his course with a similar intent of accessibility, having witnessed the same problem in a QIS course offered in a different department at his university:

\begin{iquote}
    \who{Ben}
    [The CS course's] prereqs were pretty constrictive. You needed two or three intermediate-level computer science courses ... so it wasn't really open to non-CS majors ... I tried to do the opposite -- I tried to make it available to students in any department; they just needed linear algebra.
    \timestamp{4:09 verified}
\end{iquote}

Unfortunately, perhaps because it was his first time offering the course, Ben's efforts at accessibility across majors ultimately did not prove as successful. He had only one non-physics student, whose lack of quantum mechanics background appears to have caused some challenges:

\begin{iquote}
    \who{Ben}
    I had one student who was a CS major and had never taken modern physics. I'd hoped that the quantum computing course would be equally accessible to him ...
    but this student had a little more trouble than the others.
    \timestamp{Response \#4 verified}
\end{iquote}

Ben, Carl, David, and Edwin all aim to make their course widely accessible to a diversity of backgrounds, with varying motivations and degrees of success. This focus was not universal, however. Franz, whose course is in the CS department and caters primarily to CS majors, was explicit about his target audience in the interview:

\begin{iquote}
    \who{Franz}
    It's not that I don't like math, or physics -- I love them! But this is a course for computer science students ...
    so we're very much ... focused on algorithms ... I'm focused like a laser on computer science.
    \timestamp{3:30 verified}
\end{iquote}

Franz goes on to mention that he gets some math and physics students in his class -- and even, if he recalls correctly, a linguistics student -- but emphasizes that these students are not his primary focus, and he selects content accordingly. Franz's experience appears to embody Seegerer, Michaeli, and Romeike's observation that CS experts ``agreed that a specific computer science perspective exists and is important'' within QIS education \cite{Seegerer:2021}.

Due to concerns about preparation, moreover, Albert indicated on his survey response that he had made explicit decisions to make the class more selective (though it was unclear how he accomplished this in practice):

\begin{iquote}
    \who{Albert}
    I am very selective about who can take the course, so I have very largely narrowed the possible gap between the Physics/Math major[s] and the CS majors.
    \timestamp{Survey response \#6 verified}
\end{iquote}

And despite his focus on cross-disciplinary appeal, even Carl expressed the need to balance accessibility with rigor:

\begin{iquote}
    \who{Carl}
    I didn't want to dumb it down ... I want them to be able to actually do some of the real calculations, ... see how Shor's algorithm works ...
    You know, there's just a difference between Physics for Poets and Physics 1!
    \timestamp{14:15-16:14, verified}
\end{iquote}

Edwin, in describing his accessibility-centered philosophy, cites a quote from Scott Aaronson that (perhaps inadvertently) contextualizes the trade-off we observe between accessibility and rigor: ``Quantum mechanics is really easy if you take all the physics out of it.'' \timestamp{4:26 verified}

\subsubsection{Diversity of Student Academic Levels}

Traditionally, introductory QIS courses have been largely the purvey of graduate education. As discussed below, our findings suggest, consistent with Ref.~\cite{Cervantes:2021}, that QIS education has fully expanded to undergraduate education as well.
While instructor interviews revealed tradeoffs due to differences in preparation, we heard from faculty that course design as well as the natural synergy among diverse groups of students make it possible to tailor the material for audiences of all levels. 

Each course was classified by level (first-year [FY] undergraduate, beyond-first-year [BFY] undergraduate, or graduate) according to the course's catalog number and the corresponding university definitions. Courses were classified as ``hybrid'' if they were listed for combined graduate and upper-division undergraduate audiences, either crosslisted across multiple course numbers or with a course catalog number explicitly indicating this status. Of the $N=28$ survey responses, 16 are listed (or are intended to be listed) at the undergraduate level: 2 dedicated first-year seminars and 14 BFY courses. An additional 7 courses can be classified as hybrid courses, while 5 courses are graduate-only. 
Interestingly, of the 3 courses listed or cross-listed in EE or ECE, all are graduate-exclusive courses despite our survey's bias to the contrary, suggesting that the trend toward undergraduate QIS education may be occurring primarily in physics and CS.

\begin{figure}
    \centering
       \vspace*{-1cm}
       \hspace*{-1cm}\includegraphics[width = 0.6 \textwidth]{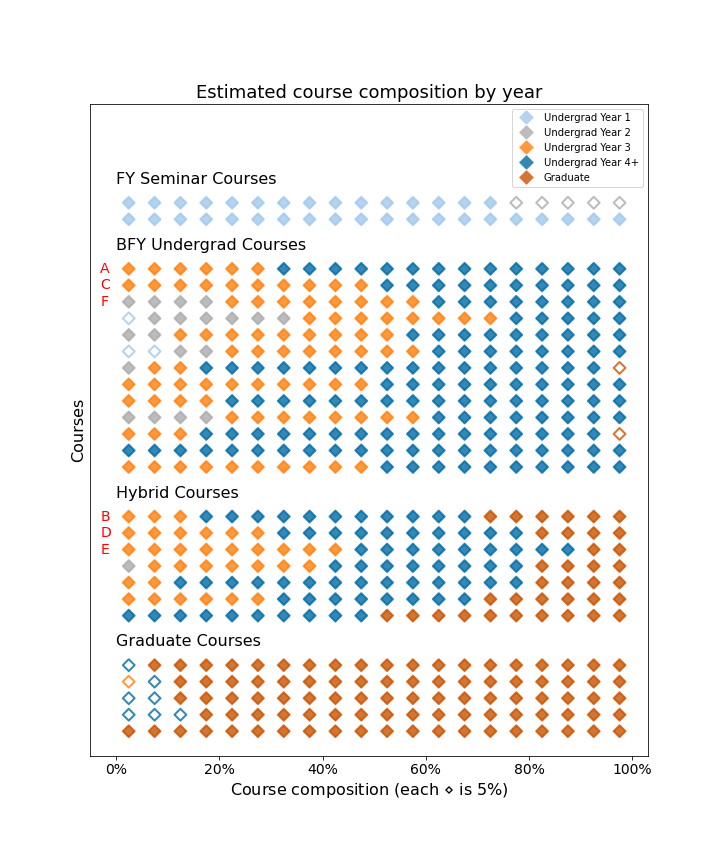}
       \vspace*{-1cm}
    \caption{Estimated enrollment profile by year for each previously-offered course ($N=27$), rounded to the nearest 5\% for each level, based on faculty responses. Courses are broken out by level classification based on university course number and show a close correspondence with this classification. Courses that are the subject of follow-up interviews are labeled A-F in red accordingly. Hollow symbols correspond to students outside the target audience listed in the course catalog.}
    \label{fig:yearprofiles}
\end{figure}

Figure \ref{fig:yearprofiles} shows the distribution of student years for each course, broken down by level classification. With the exception of the two first-year seminar courses, no course has greater than 10\% freshmen, and most have similarly low levels of sophomores; the bias toward 3rd year and higher might be understood as a reflection of course prerequisites required by many introductory QIS courses as well as the audience toward which these courses are marketed. Also, Fig.~\ref{fig:yearprofiles} shows a close correlation between the actual course composition and the course level classification.

Similar to choices about which major or academic background(s) of students to target, decisions about what level(s) of students to target in a course also are subject to a series of tradeoffs. For example, one theme that emerged from a free-response survey question regarding student difficulties and gaps in understanding was that understanding quantum hardware and devices requires a depth accessible only to fourth-year undergraduates or graduate students, but that the class can be made more accessible for younger students by focusing on algorithms:

\begin{iquote}
\who{Respondent \#7}
If one focuses on quantum algorithms, there is relatively little preparation needed ... Any discussion on qubit implementations requires a lot of physics -- quantum mechanics, condensed matter, etc. -- and is only appropriate for seniors, if at all.
\timestamp{Response \#7 verified}
\end{iquote}

Respondent \#7 was teaching at a primarily-undergraduate institution and thus made the decision to exclude most content on qubit implementations to match the student audience. Respondent \#11 cited similar reasons for opting to teach a graduate-only course so that such material could be included. This challenge was particularly salient for FY seminar courses, as seen in Respondent \#27's survey response:

\begin{iquote}
\who{Respondent \#27}
 I do not assume any physics background, and this poses a limitation in explaining the hardware of various quantum computing platforms. Thus, I end up focusing primarily on theoretical aspects of quantum information, and giving only hand-waving explanations of the hardware involved ... 
 \timestamp{Response \#27 verified}
\end{iquote}

Material that may be of interest to graduate students, particularly concerning hardware, can be inaccessible to undergraduates. One solution, adopted by Carl's institution, is to offer separate undergraduate and graduate courses (Carl's response focuses only on his undergraduate course):

\begin{iquote}
\who{Carl}
Well, one impetus for me [in developing the undergraduate course] was that occasionally undergraduates have taken ... the graduate level class. And they do very poorly ... 
it goes too fast.
\timestamp{3:45 verified}
\end{iquote}

Alternatively, we can look to hybrid undergraduate-graduate courses to see how instructors manage to bridge this gap.
Ben designed a course targeting both undergraduate and graduate student levels by providing supplementary reading and homework for graduate students, particularly on physical realizations of quantum gates:

\begin{iquote}
\who{Ben}
For the undergrads ... I think we may not have mentioned the Schrodinger equation! Whereas for the grad students, I had them read the chapters ... [on] development of quantum gates from Hamiltonians.
\timestamp{6:25 verified}
\end{iquote}

David, on the other hand, argues that it is not particularly difficult to design a course meeting the needs of both undergraduate and graduate students. He reframes the diversity of student backgrounds -- in discipline as well as level -- as an instructional asset, with the diverse intellectual backgrounds of students combining to build a broad knowledge base in the fundamentals of quantum mechanics:

\begin{iquote}
\who{David}
It just -- it didn't seem unreasonable to be able to teach it at a level where a physics undergrad who ... had taken the advanced quantum course or was concurrently taking the advanced quantum course. And ... our graduate students, who had had more exposure. And in engineering who'd had more exposure to quantum mechanics already; they could all sort of reasonably take it at the same level, so -- and it seemed to work.
\timestamp{11:30 verified}
\end{iquote}

\subsubsection{Prerequisites Vary Across Courses}

In addition to the level and majors of students, we also asked survey respondents to report any prerequisites required for enrollment in their course. We find that the prerequisite courses in math, physics, and CS (if any) vary substantially across courses. We also find that course design is impacted by the inevitable trade-off between accessibility and the benefits of prior knowledge, with courses requiring fewer prerequisites having to cover more non-QIS background material.

Courses were classified as a prerequisite if they were required or strongly recommended for enrollment in the course, even if the instructor grants occasional case-by-case exemptions. 
Math methods courses in physics or CS courses were classified according to their relevant content areas.

\begin{figure}
    \centering
    \begin{subfigure}{0.5 \textwidth}
        \hspace*{-1cm}
        \includegraphics[width =    1.2 \textwidth]{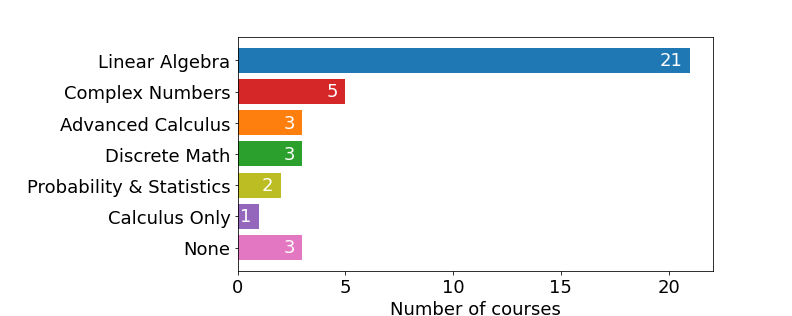}
        \caption{\textbf{All} required math preparation at or beyond calculus level; some courses require more than one math prerequisite. Discrete math includes number theory and/or related topics. Advanced calculus includes multivariable calculus, differential equations, and/or related topics.}
        \label{subfig:prepmath}
    \end{subfigure}
    \begin{subfigure}{0.5\textwidth}
        \hspace*{-1cm}
        \includegraphics[width = 1.2 \textwidth]{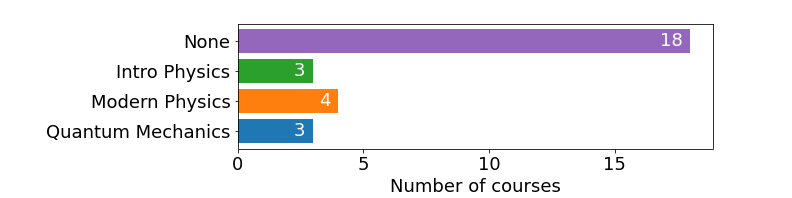}
        \caption{\textbf{Highest} required physics preparation beyond high school level. Intro or physics or quantum mechanics prerequisite may be 1 or 2 terms depending on the course and institutional calendar.}
        \label{subfig:prepphysics}
    \end{subfigure}
    \begin{subfigure}{0.5\textwidth}
        \hspace*{-1cm}
        \includegraphics[width =1.2 \textwidth]{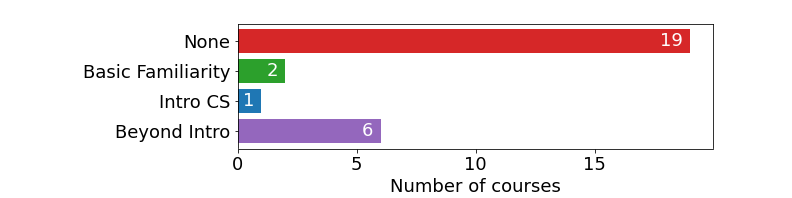}
        \caption{\textbf{Highest} required computer science preparation. Basic familiarity indicates informal experience with programming but not necessarily in a classroom setting. Beyond intro courses refers to any courses following the first semester of introductory computer science for technical majors, and can include topics such as algorithms or data structures.}
        \label{subfig:prepcs}
    \end{subfigure}
    \caption{Required prerequisite courses for each introduction to QIS course, as reported by the $N=28$ instructors (we include Respondent \#23 since he has already decided upon prerequisites for his course). Note that \ref{subfig:prepphysics} and \ref{subfig:prepcs} list only the highest required prerequisite for each course. By contrast, \ref{subfig:prepmath} lists all required courses above calculus, since math course sequences are not typically standardized across universities beyond calculus.} 
    \label{fig:prereqs}
\end{figure}

Figure \ref{fig:prereqs} shows the prerequisites given by the instructor for each of the 28 intro to QIS courses surveyed, broken out by topic (physics, CS, or mathematics). Prerequisites in other disciplines, such as engineering, were not asked about; among the courses that submitted optional syllabi, however, none listed prerequisite courses in such disciplines. Respondent \#23's course was included since the instructor had already determined the prerequisites he intended to require, even though the course had not actually been run yet.
Note that most courses (19/28, 68\%) require no CS prerequisite, and a similar fraction (18/28, 64\%) require no physics prerequisite. Linear algebra is the most common prerequisite, required by 21 of 28 courses (75\%), though many instructors allow students to select from multiple courses with linear algebra content. 

One frequently reported challenge was the tension between the benefits and impracticality of requiring prerequisite courses. 
In the following quote, Carl discusses his teaching team's decision not to require modern physics or discrete math as prerequisite courses, arguing that to require either would make the course inaccessible to a large portion of the intended audience for the course:

\begin{iquote}
\who{Carl}
When we were designing, we had a lot of discussion, like, it would really be better if we had [modern physics] as a prerequisite. But that limits us to physics students. It would be much easier if we had ... the discrete math class the computer scientists take, but essentially no physics students take that.
\timestamp{21:25 verified}
\end{iquote}

Franz describes a similar dilemma in the context of his course targeting CS students at a small liberal arts college:

\begin{iquote}
\who{Franz}
Even though a course in quantum computing really could benefit from prerequisites in, like, classical circuit design, quantum theory, like more advanced linear algebra, more advanced like probabilistic algorithms, complexity theory, Fourier analysis -- there are like all these things which the ideal student would have, but like, I don't dare assume any of that, right?
\timestamp{5:04 verified}
\end{iquote}

Given the already small pool of students eligible to take the course, we asked Franz to elaborate on why unlike most instructors he required a second semester of CS coursework. His response was particularly informative:

\begin{iquote}
\who{Franz}
Actually, I doubt that we use really any material from the second course. So then why do I require it? 
... I felt like students who come out of our very first intro CS course, yes they've seen some computer science, but they're still just not good at it, basically ... It's this kind of vague, I guess, argument of maturity.
\timestamp{17:55 verified}
\end{iquote}

The authors view Franz's theme of maturity, echoed in similar language in the context of mathematics by Respondent \#20, as suggesting that prior content exposure may be less important for students in intro QIS courses than mathematical and algorithmic sophistication. 

Surprisingly, some instructors found that additional prerequisites, even when directly relevant to the content, are not actually helpful for student learning. Edwin has found no benefit in students completing a semester of undergraduate quantum mechanics, since the wave functions-first curriculum used at his institution does little to prepare students for the discrete vector spaces used in QIS. David related:

\begin{iquote}
\who{David}
We found that ... how much quantum mechanics the students had previously been exposed to was not a very good indicator [of] how much trouble, or how well they did, in the course.
\timestamp{13:37 verified}
\end{iquote}

Respondent \#20 even opted against a prerequisite in linear algebra -- a prerequisite required by 75\% of surveyed courses -- having found, according to a statement in his syllabus, no statistically significant difference in student performance. In light of Respondent \#20's aforementioned emphasis on mathematical maturity -- indeed, he offers optional out-of-class linear algebra review sessions to mitigate gaps in content coverage -- we hypothesize that the disparity in instructor views on prerequisites reflects deeper disagreement on the role of prerequisite courses: are they prescribed to cover content or to build maturity?

In discussing the diversity of student backgrounds, particularly in the context of cross-disciplinary courses, a common theme emerged regarding the effects of diverse student backgrounds on the course as a whole. Some groups of students -- usually physics or CS majors -- would succeed more easily in one section of the course than in another. Additionally, student difficulties varied across majors.
David reported that the diversity of students' math backgrounds led to issues with boredom among students who already had strong math backgrounds. Likewise, some instructors discussed asymmetries between the backgrounds of physics and CS students, though this was not a universal experience. Carl reported that CS students tended to struggle with quantum measurement but had a much easier time with combinatorics than physics students. Edwin likewise noted that physics students tended to find the number theory and group theory associated with quantum algorithms particularly difficult. David reported that engineering students were much more likely than physics students to be comfortable with complex vectors. Regarding his graduate course, one survey respondent summarized:

\begin{iquote}
\who{Respondent \#19}
Each major perform[s] better at different parts of the course.
\timestamp{Response \#19 verified}
\end{iquote}

Finally, multiple instructors reported needing to make adjustments to their teaching practices to accommodate the backgrounds of their students. Carl reported needing to make changes to the lecture slides due to incorrect assumptions about CS students' math backgrounds. As we saw earlier, Ben found it useful to assign additional readings to graduate students to better align the level of the course to students' background. Franz perhaps summarizes best the trade-offs associated with gaps in students' backgrounds by noting that as a result, much of the material in the course won't be directly about quantum computing:

\begin{iquote}
\who{Franz}
I have to explain how to factor integers, even though the jump from period finding [quantum subroutine in Shor's algorithm] to factoring integers has nothing to do with quantum computing ... Actually throughout the course, there's a fair amount of material that's just not quantum computing. And so that's the kind of trade-off you make.
\timestamp{22:19 verified}
\end{iquote}

\subsection{Diversity of Instructor Backgrounds and Its Effect on Instruction}

Quantum information science is a broadly interdisciplinary field, encompassing areas as disparate as physics, mathematics, computer science, and electrical engineering. As a result, relatively few faculty are experts in the full range of topics within QIS. Some, such as Albert and Franz, stumbled into teaching these courses with essentially no QIS background whatsoever. Moreover, though a preponderance of instructors (17) were physics faculty, the faculty we surveyed hail from a diversity  of departments: 6 instructors from a computer science department, 3 from an electrical engineering or ECE department, 1 from a math department, and 1 from an explicitly interdisciplinary research division. The subsequent sections explore three ways this diversity of instructor backgrounds affects introductory QIS instruction.

\subsubsection{Variety of Goals for Student Learning}

Consistent with their own diverse experiences and preparation, we observed a broad diversity of learning goals and intentions across the 6 faculty interviewees. We
focus on faculty's varied perspectives on the relative importance of two topics: fundamental physics and quantum programming.

Albert was emphatic that his primary goal was to build excitement and intuition for the underlying mathematical formalism and physics of quantum mechanics:

\begin{iquote}
\who{Albert}
I don't give a damn about quantum computation, truth be told. I care about them learning quantum mechanics. That's all I really care about.
\timestamp{1:03:49 verified}
\end{iquote}

Albert's course accordingly
eschews most quantum algorithms in favor of topics such as Bell's inequality, and even focuses on the quantization of the electric field in order to physically realize a CNOT gate (a topic very much tangential to the scope of QIS as defined by Ref.~\cite{Seegerer:2021}). 
David shares with Albert a strong focus on the underlying math and physics:
\begin{iquote}
\who{David}
The most basic thing is we'd like them to get -- come away with a better understanding and ability to use the basic formalism of quantum mechanics, ... the basic setup in complex vector spaces. 
\timestamp{30:21 verified}
\end{iquote}

By contrast -- and this connects with his choice to target computer science students as an audience -- Franz is as adamant about the relative \textit{unimportance} of fundamental physics to his course as Albert and David are to theirs: 

\begin{iquote}
\who{Franz}
So I really ... in some sense try to teach as little math and physics as possible. \timestamp{3:30 verified}
\end{iquote}

Ben, Carl, David, and Franz all identified the ability to interpret and execute a quantum circuit diagram, computing the final state from the initial state, as an essential learning goal for their course. However, the natural extension of this idea -- actually programming quantum computers -- ranged in importance from essential to irrelevant in the views of respective interviewees. For example, Ben specifically wants his students to be able to translate between circuit diagrams and programming a real quantum computer on the cloud:

\begin{iquote}
\who{Ben}
Basically just looking at a quantum circuit diagram, understanding the Hadamards, the CNOTs, the other [quantum gates] -- just understanding how to calculate what the final state is based on the [initial] state. And understanding how to run it on IBM Quantum on real remote devices and get data.
\timestamp{8:01 verified}
\end{iquote}

Edwin likewise considers the programming of a quantum computer an essential skill for the course. Franz doesn't have his students extensively program quantum computers directly, but requires his students to code a classical simulation of a quantum computer. Meanwhile, Carl dispenses with programming entirely, arguing that it merely distracts from the fundamental focus on linear algebra and quantum gates:

\begin{iquote}
\who{Carl}
I think [programming is] very much beside the point of the topic of quantum computing. I think programming for quantum computers is 95\% the same as programming for classical computers, and ... teaching someone how to program classical computers just kind of isn't necessary or of value when we're trying to help people understand how quantum computers work.
\timestamp{38:45 verified}
\end{iquote}

Interestingly, none of the instructors have as a goal for students the ability to develop their own quantum algorithms. Franz explains that this goal would be impossible for his students because ``they [quantum algorithms]'re too hard.'' In his survey response, Franz elaborates:

\begin{iquote}
\who{Franz}
(A) Most of the intuition from classical CS doesn't apply to the design of quantum algorithms, so we have to develop an entirely new intuition about how to design quantum algorithms, but (B) there are so few quantum algorithms (especially accessible, motivated ones) that we can't really learn by example.
\timestamp{Response \#8 verified}
\end{iquote}

\subsubsection{Opportunities for Collaborative Course Development}

Given the interdisciplinary nature of QIS as a field, it is not surprising that some classes are co-taught by multiple instructors. David developed his course as part of a team of four, alongside one other physicist, one electrical engineer, and a former faculty member in electrical engineering and computer science whose research he describes as mostly mathematics. Additional input and guest lectures were provided by the wife of one of his collaborators, herself a professor in computer science and mathematics with a research focus on quantum measures of entropy. Likewise, Carl co-developed his course with two colleagues, one a fellow physicist and the other a computer science theorist whose graduate work involved quantum computing.

For David and Carl, having a diverse team was particularly useful given the ability to draw on the expertise of colleagues:

\begin{iquote}
\who{David}
[Two collaborators] were doing actual research in, in questions having to do with quantum information. So they were experts -- I mean, not so much in the whole algorithm-type thing, but you know ... certainly I learned an awful lot from them.
\timestamp{56:05 verified}
\end{iquote}

For David, who has relatively little background in QIS, working with a team has helped him solidify his own background. For Carl, himself an expert, having multiple disciplinary perspectives is a way to help students see the material from multiple angles. In both courses, there is the additional benefit that each instructor may teach the material that is most relevant to their own personal expertise; for instance, Carl's lectures might focus on fundamental quantum mechanics while his computer science co-instructor would focus more on algorithms.

\subsubsection{Primary Literature Not Always Accessible to Instructors}

Relatively few of the instructors we surveyed and interviewed identified as experts in QIS. As a result, Albert, Ben, David, and Franz all expressed, in one context or another, the importance of proactive self-education in QIS education. 
Franz identified a particular challenge pertaining to self-education in the context of his background as a nonexpert. As a rapidly evolving discipline, teaching an up-to-date QIS course requires a knowledge of current research, yet primary literature published in journals is often not accessible to those without a background in QIS. As he writes:

\begin{iquote}
\who{Franz}
It's not easy for me (as a non-specialist) to read the primary literature, and it's really difficult to make it accessible to my undergraduate CS audience.
\timestamp{Response \#8 verified}
\end{iquote}

He suggests ``digestible treatments of recent research developments'' as a tool he as an educator would like, adding:

\begin{iquote}
\who{Franz}
It seems to me that QIS is young enough that a lot of people who are teaching it are non-specialists ... I worry that my curriculum is 20 years out of date and I don't even know it.
\timestamp{Response \#8 verified}
\end{iquote}

Franz's comments suggest a broader need in the QIS community for more accessible scientific communication of novel results, aimed at a non-specialist but technical audience.

\subsection{Particular Challenges Reported by QIS Instructors}

Survey respondents were asked to report any particular challenges they observed in teaching introduction to QIS courses, encompassing both observed student difficulties and challenges facing individual instructors.
We highlight two particularly consistent themes emergent from our analysis of the survey responses and subsequent follow-up interviews: student difficulties with mathematics, especially linear algebra, and concerns with available textbook quality for introduction to QIS courses particularly at the undergraduate level.

\subsubsection{Student Difficulties with Mathematics, Especially Linear Algebra}\label{sec:linearalgebra}

Survey respondents were asked to list student difficulties they observed in their course. We observed that student difficulties with mathematics were by far the most frequently reported category of student difficulty, being reported by 17 of 27 instructors who had previously taught at least one iteration of their course. Fully 14 instructors reported student difficulties with linear algebra.
We find that bra-ket notation \cite{Singh:2013,Marshman:2015}, tensor products, and the spectral theorem\footnote{The spectral theorem, foundational to quantum measurement, states that a Hermitian matrix (the complex generalization of a real symmetric matrix) can be diagonalized in a basis of orthonormal eigenvectors with real eigenvalues.} emerge as subtopics of linear algebra that are particular sites of instructor-reported difficulty.

Free-response survey answers were coded by topic, which were then grouped into broader emergent areas encompassing the full range of coded responses (mathematics, quantum mechanics, computer science/coding, other). We expect that this methodology will tend to underestimate the prevalence of specific student difficulties, since null responses do not necessarily imply the absence of a particular student difficulty.

Table~\ref{tab:difficulties} shows the student difficulties reported by instructors, as broken down by category. Notice that difficulties with mathematics dominate those of other categories, with 17 of 27 (63\%) of surveyed instructors reporting one or more difficulties with mathematics. A smaller but significant fraction (37\%) reported difficulties with quantum mechanics. Only one instructor noted a student difficulty with programming or related computer science topics, perhaps reflecting Carl's sentiment that ``physics students know more about computers than computer science students know about physics'' and the observation that only a subset of instructors emphasize programming.

\begin{table}[]
    \centering
    \begin{tabular}{p{2.6cm} c p{2.35cm} c}
        \hline \hline
        \thead{Student Difficulty} & \thead{Count} & \thead{Math Subtopic} & \thead{Count}\\
        \Xhline{2\arrayrulewidth}
        Math (any) & 17 & Linear algebra & 14\\
        Quantum mechanics & 10 & Complex numbers & 6\\
        Programming & 1 & Group theory & 3\\
        Other & 3 & Prob/stat/combinatorics & 3\\
        & & Calculus & 2\\
        & & Math (general) & 4 \\
        \Xhline{2\arrayrulewidth}
        \thead{Any difficulties\\reported} & 23 & & 17\\
        \hline \hline
    \end{tabular}
    \caption{Categorized student difficulties as reported by the $N=27$ instructors who had previously taught at least one iteration of their course. Note that 23 of the 27 faculty listed at least one student difficulty, and 17 of 27 listed at least one difficulty in the domain of mathematics.}
    \label{tab:difficulties}
\end{table}

The right-hand column of Table~\ref{tab:difficulties} further breaks down the student difficulties in mathematics by subtopic. Note that 14 instructors cited one or more student difficulties with linear algebra, representing 52\% of the survey respondents who had previously taught at least one iteration of their course and an overwhelming 82\% majority of those respondents who reported at least one difficulty with mathematics. 
The large number of responses related to student difficulties with linear algebra is perhaps unsurprising, given the centrality of linear algebra in QIS. As Carl explains:

\begin{iquote}
\who{Carl}
Quantum computing to zeroth order is linear algebra.
\timestamp{42:49 verified}
\end{iquote}

Analysis of particular survey and interview responses provides greater insight into faculty perceptions of the nature of the student difficulties with linear algebra.
Although students may enter the course able to perform matrix mechanics and manipulations, they lack the deeper understanding of, and intuition for, linear algebra necessary to apply the material to the abstract problems and vector spaces encountered in QIS:

\begin{iquote}
\who{Franz}
If we phrase linear algebra as like, can you multiply this $2^N$ matrix by this $2^N$ matrix ... if we make it that concrete, I think my students have basically no problem with it -- none. They start to get a problem when things are not written out that way, but are instead written out as like tensor products of states, right? Because that then is like a level of abstraction.
\timestamp{39:52 verified}
\end{iquote}

Franz further related that bra-ket (Dirac) notation was particularly challenging for his computer science students:

\begin{iquote}
\who{Franz}
I mean, arguably, it's notation ... there's like kets flying around all over the place!
\timestamp{41:17 verified}
\end{iquote}

The student difficulties with bra-ket notation are consistent with Respondent \#10's observation that students came into the course lacking \textit{fluency} in linear algebra even if they were familiar with basic concepts. Bra-ket notation abstracts away much of the concreteness of matrix notation in favor of the underlying, basis-independent mathematical structure. Students who are familiar with matrix mechanics but lack the fluency needed to abstract beyond rote calculations would likely struggle to learn bra-ket notation precisely for this reason. For instance, David finds that one particular student difficulty with bra-ket notation comes in forgetting that the symbol inside the bracket is merely an abstract label, giving the example that students will sometimes write $2\bra{\psi}$ as $\bra{2\psi}$. David also finds that students struggle with applying gates to the correct qubit when translating between a quantum circuit diagram and bra-ket notation.

Both Carl and David emphasized tensor products as a particular difficulty for their students:

\begin{iquote}
\who{David}
A lot of our students end up at the core still confused about tensor products.
\timestamp{1:10:55 verified}
\end{iquote}

A major issue seems to be that linear algebra courses at the university level do not prepare students well for quantum computing; the topics and skills covered are simply too different. In fact, as Carl and Edwin point out, linear algebra concepts such as tensor product spaces and the spectral theorem, which are considered foundational to QIS, receive little to no attention in undergraduate linear algebra courses:

\begin{iquote}
\who{Carl}
So linear algebra classes, it turns out, do not teach you about tensor products usually, and even don't talk about vector spaces much.
\timestamp{2:32 verified}
\end{iquote}

Of course, undergraduate introductory linear algebra courses are intended for applicability to a large range of majors; it is not  realistic to expect these courses to equip incoming students with all the mathematical sophistication they require for success in a QIS course. Carl found that an intensive linear algebra review at the beginning of the course was effective in avoiding some of the pitfalls with student difficulties he experienced the first time he taught the course. And while general-audience linear algebra courses may never suit the needs of QIS students, discipline-specific math methods classes hold promise: Both David in physics and Carl's co-instructor in CS have had success adapting their respective disciplines' math methods classes to better emphasize the necessary mathematical skills and concepts for QIS.

\subsubsection{Concerns About Textbook Quality for Undergraduate QIS}

Survey respondents were asked to report the textbook(s) required for their course, if any. 
There was relatively little consensus around any one particular textbook, and faculty perceived existing QIS textbooks as being poorly suited for undergraduate courses.

Of the 27 courses previously offered -- Respondent \#23 was still in the process of selecting a textbook at the time of the survey -- a total of 14 unique textbooks (not counting separate editions of the same textbook) were reported as required by one or more courses. Of these, only 4 textbooks were required by more than one course, and only 2 textbooks were required by more than 2 courses: \textit{Quantum Computer Science: An Introduction}, by N. David Mermin (2003), assigned by 7 faculty, and \textit{Quantum Computation and Quantum Information}, by Michael A. Nielsen and Isaac L. Chuang (2011), assigned by 6 faculty. Tellingly, of the 27 courses, 9 courses assigned no textbook at all -- in some cases drawing on extensive self-written course notes -- while 6 courses required more than one textbook. David confirms that his primary motivation for writing his own course notes was textbook quality:

\begin{iquote}
\who{David}
All of this grew out of our dissatisfaction with the textbooks.
\timestamp{49:54 verified}
\end{iquote}

Complaints about the quality and suitability of existing textbooks were a prominent theme across survey responses, even though faculty were not specifically prompted to consider the quality of textbooks. A common theme was that existing textbooks are not appropriate for teaching upper-division undergraduate courses, either being ``too mathematical'' for undergraduates (in the words of Respondent \#23) or not rigorous enough for the goals of the course:

\begin{iquote}
\who{Respondent \#16}
All the texts I've looked at are either too low level or too high level.
\timestamp{Response \#16 verified}
\end{iquote}

Even among the faculty who utilized Mermin and/or Nielsen \& Chuang, the consensus was that these textbooks were not optimal for undergraduates. In general, faculty perceived Mermin as too mathematically-dense and Nielsen \& Chuang as requiring too much physics for undergraduates.
As QIS education matures, the demand certainly exists for textbooks informed by education research and with an emphasis on current topics. However, the wide range of instructor goals suggests no textbook is likely to suit the needs of every course.

\section{Discussion and Conclusions}

One possible outcome of this study would have been
to define a canonical introductory QIS course.
Instead, we find that each course is intended by its instructor to tell a slightly different story about QIS even if the course topics are ostensibly similar.
A freshman seminar and a graduate quantum information/quantum optics course clearly wouldn't be expected to have the same structure and learning goals. What we learned is that neither do, say, two BFY undergraduate courses in computer science departments (e.g.\ Albert and Franz) nor two hybrid graduate-undergraduate courses both crosslisted in computer science and physics (e.g.\ David and Edwin). 
At some level, the five core concepts identified by Seegerer \textit{et al.} \cite{Seegerer:2021} did appear to emerge to some degree across all courses, but instructors' treatment of these core concepts varied dramatically from course to course. ``Quantum algorithms'' was a concept covered by all six of the interviewed instructors, but it ranges from the central theme in Franz's course to a peripheral one in Albert's, for instance.

With the growing prominence of QIS as a discipline and the recognized need to develop a quantum-ready workforce \cite{Aiello:2021, Fox:2020}, there will undoubtedly be pressure for QIS instruction to standardize. Greater agreement on the topics to cover, and more fundamentally the learning goals to emphasize within QIS education, would facilitate the development of better textbooks and curricular materials. Greater standardization would also promote involvement of education researchers in QIS education as called for by Aiello \textit{et al.} \cite{Aiello:2021}.

Yet it is important to make sure that canonicalization does not come at the expense of the discipline's interdisciplinary perspective, which is arguably among the greatest strengths of QIS as a research area. Our work here suggests that diversity within and among QIS courses can likewise be a valuable asset worth preserving: though some instructional \textit{methodologies} will certainly be more effective than others, the variety of orientations to and goals for QIS instruction seen across the courses profiled in this paper will help preserve the diversity of thought that has helped the quantum workforce thrive.

\begin{figure}
    \centering
    \hspace*{-0.5cm}\includegraphics[scale=0.33]{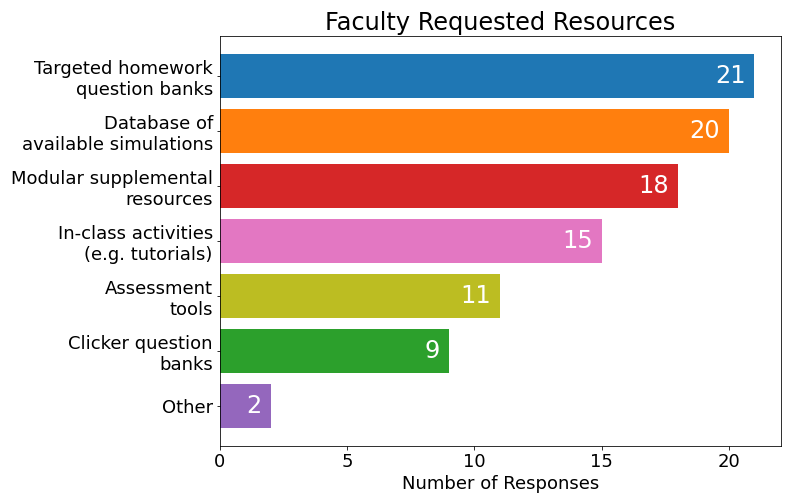}
    \caption{Number of survey respondents expressing interest in implementing various proposed PER curricular instruments in their classrooms. ($N=28$, instructors allowed to select any number of responses)}
    \label{fig:resources}
\end{figure}

Our findings demonstrate ample room for DBER groups to become actively involved in quantum computing education through development of curricular and assessment tools. Figure \ref{fig:resources} shows moderate-to-high faculty interest among survey respondents in research-based educational materials that DBER researchers could develop. Unsurprisingly, support was strongest for resources easily deployed within traditional classroom settings, such as targeted homework question banks. Interest in assessment tools, as well as active-learning resources such as tutorials and clicker questions, were lower (predictably, given that these innovations are lesser-known) but still substantial with approximately 30 to 50\% interest. 
Support for assessment and active-learning tools grew among interviewees once the benefits of these instruments were explained, suggesting that interventions targeting these areas still have ample space to succeed as long as the benefits are clearly promoted to instructors. David, for instance, 
realized an assessment might help him understand his students' struggles with tensor products. Franz likewise became interested in assessment tools when the interviewer explained to him that such tools could be multiple-choice, thereby overcoming his concern that an assessment would require too much effort to administer.

\subsection{Directions for Further Research}

QIS education remains a largely unexplored frontier in PER and related discipline-specific education research fields, and presents a rare opportunity for education researchers to be involved proactively in the development of an emerging but rapidly-growing field of coursework as called for by Aiello \textit{et al.} \cite{Aiello:2021}. As such, more research is necessary on every front, from understanding student reasoning in QIS contexts to curricular and assessment development. This work investigates QIS courses at US institutions from the instructor's perspective, and our team's prior work examined the same topic by catalog listing \cite{Cervantes:2021}; student interviews across a similarly broad cross-section of courses would be useful for contextualizing the findings reported here.
Likewise, all interviewees and a sizeable majority of survey respondents for this work identified as white and male, reinforcing the need for additional scholarship on the experiences and perspectives of marginalized groups in QIS education. Concerns of marginalized student populations are particularly important to highlight given the observed disparities between undergraduate QIS course offerings between private and public universities and between minority-serving and non-minority-serving institutions \cite{Cervantes:2021}, including more detailed analysis of how QIS education intersects with issues of diversity, equity, inclusion, and justice (DEIJ).

Though faculty's survey and interview responses revealed a variety of goals regarding student learning in the course, few instructors commented on the specific jobs they wish to prepare students for. 
An important question for further work is what careers (if any) faculty are envisioning they are preparing their students for. There may be a disconnect between industry needs for workforce development \cite{Fox:2020, Aiello:2021, Hughes:2022} and the goals of existing QIS coursework, a divide that perhaps gets at the heart of the tension we observe between (in Carl's words) accessibility and the risk of ``dumb[ing] it down.''

Finally, the state of QIS education remains rapidly in flux, with a number of trends likely to have dramatic effects on the state of QIS education in the coming years. The NSF's Q-12 initiative aims to bring quantum education to the K-12 classroom, thereby potentially changing the background and demographics of tomorrow's QIS students at the university level. And a number of US institutions -- including the University of California Los Angeles, the University of Southern California, the University of Wisconsin Madison, Colorado School of Mines, and the University of Rhode Island -- are joining a global trend in piloting professional master's programs in QIS or closely-related fields. These programs represent ideal testbeds for discipline-based education researchers to be involved in shaping the future of quantum computing education. At the same time, the rapid growth in QIS education means that follow-up studies
will be imperative to remain apprised of trends as QIS coursework continues to evolve.

\section{Acknowledgments}

Special thanks to Bianca Cervantes for her work identifying candidate courses to survey, Martin Laforest for his feedback on the faculty survey design, and to our faculty respondents and interviewees who made this effort possible. This research is funded by the University of Colorado Boulder Department of Physics, the NSF Graduate Research Fellowship Program, and NSF Grants No.'s 2012147 and 2011958.


\end{document}